\documentclass[structabstract]{aa}  
\usepackage{graphicx}
\usepackage{natbib}
\bibpunct{(}{)}{;}{a}{}{,}

\usepackage{txfonts}
\begin{document}
   \title{The puzzling temporally variable optical and X-ray afterglow of GRB 101024A}

   \author{
B. Gendre\inst{1}					\and
G. Stratta\inst{1}				\and
M. Laas-Bourez\inst{2}						\and
D. M. Coward\inst{3}					\and
A. Klotz\inst{4}$^, $\inst{5}	\and
S. Cutini\inst{1}					\and
M. Bo\"er\inst{5} \and
C. Stockdale\inst{6}
          }

   \institute{
ASI Science Data Center, via Galileo Galilei, 00044 Frascati (RM), Italy
              \email{bruce.gendre@asdc.asi.it}
         \and
University of Western Australia, School of Physics/ICRAR, Crawley, W.A. 6009, Australia
				 \and
University of Western Australia, School of Physics, Crawley, W.A. 6009, Australia
         \and
Institut de Recherche en Astrophysique et Planetologie, 9 avenue du colonel Roche, 31028 Toulouse, France
         \and 
Observatoire de Haute Provence (CNRS), 04780 Saint Michel l'Observatoire, France
				\and
Centre for Backyard Astrophysics, Churchill  Victoria  3842, Australia
}

   \date{Received ; accepted }

% \abstract{}{}{}{}{} 
% 5 {} token are mandatory
 
  \abstract
   {}
   { To present the optical observations of the afterglow of GRB 101024A and to try to reconcile these observations with the X-ray afterglow data of GRB 101024A using current afterglow models}
   {We employ early optical observations using the Zadko Telescope combined with X-ray data and compare with the reverse shock/forward shock model.}
   {The early optical light curve reveals a very unusual steep decay index of $\alpha\sim5$. This is followed by a flattening and possibly a plateau phase coincident with a similar feature in the X-ray. We discuss these observations in the framework of the standard reverse shock/forward shock model and energy injection.We note that the plateau phase might also be the signature of the formation of a new magnetar. }
   {}


   \maketitle
%
%________________________________________________________________

\section{Introduction}

Gamma-Ray Bursts (GRBs) are the most powerful explosions in the Universe since the Big Bang \citep[see e.g.][for a review]{mes06}. GRBs are thought to be the signature of the formation of a black hole via the collapse of a massive star in a hypernova or the merging of two compact objects \citep{mes06}. It has also been proposed that newborn magnetars could also manifest as GRBs \citep{dai98}. The gamma-ray burst is observed for the first seconds of the event, followed by an afterglow at longer wavelengths. However, in a few cases, coincident optical and gamma ray burst emission has been observed \citep[see][for a recent global view of the early part of the optical light curve]{gen09}, and for this reason it is more correct to call this initial part of the phenomenon the {\it prompt phase} of the burst.

In the most commonly accepted model, the radiation mechanism is supposed to be synchrotron radiation from an expanding fireball \citep{ree92, mes97, pan98}. In the first model, all the parameters of the fireball are fixed at start, and the observed light curve shape should be very simple \citep{sar98,sar99}. {\em Swift} has however shown that this view was too simplistics: in the X-ray part alone, the afterglow shows unexpected features such as flares, a plateau phase, chromatic breaks and the absence of an achromatic late break. These features are not explained by this model \citep[see e.g.][]{gen10}. Among these features, the plateau phase is one of the most complicated to explain. It is thought to be due to late energy injection in the fireball \citep[e.g.][]{pan06} or the emergence of the afterglow hidden during the early part by the end of the prompt phase \citep{wil07}, see \citet{zha07} for a complete review. These interpretations lack the support of multi-wavelength observations. In almost all cases, they are used to explain the observed light curve in a certain band without knowledge of other bands (and/or the spectral properties): as emphasized above, the gamma-ray burst phenomenon is a multi-wavelength phenomenon. 

In this paper, we present the observations of \object{GRB 101024A} observed by {\em Swift} and several robotic telescopes. We highlight the properties of GRB 101024A in Sections \ref{sec_burst} and \ref{sec_analyze}; and discuss the early optical phase in Section \ref{sec_sample} and  the plateau nature in Section \ref{sec_discu} before concluding. All errors are given at the 90\% confidence level. %When needed, we used a simple flat Universe with $\Omega_\Lambda=0.3$.

\section{GRB 101024A}
\label{sec_burst}

\object{GRB 101024A} was detected by the {\em Swift} satellite \citep{ghe04} at 11:39:29.8 UT (hereafter noted $T_0$) on 2010 October 24 \citep{dep10}, and also observed by Fermi \citep{mcb10}. We analyzed the BAT and Fermi prompt data, and obtained a duration of $T_{90} = 18.7 \pm 0.5$s, and a best fit spectral parameter of $\Gamma = 0.7 \pm 0.4$, $E_{cut} = 46^{+24}_{-13}$ keV using a cut-off power law. The XRT position of the burst is RA = 04$^h$ 26$^m$ 1.51$^s$ DEC = $-77\deg 15\arcmin 55.4\arcsec$ with an uncertainty of 1.8$\arcsec$ \citep{bea10}. Because of its position in the sky and the time at which the burst occurred, only a few robotic telescopes responded to the alert in optical. The Zadko telescope \citep{cow09} was the first to observe the optical afterglow. \citet{laa10a} reported the observation of a fading afterglow. The afterglow was also observed by GRAS06 \citep{hen10} and by AAVSO International High Energy Network \citep{sto10}. Finally, the UVOT on board {\em Swift} reported observations of the afterglow in the U and V bands \citep{dep10}.

The afterglow decayed very quickly: about 20 hours after the trigger it was too faint to be detected by the X-Shooter mounted on the VLT-2 unit (D'Elia 2010, private communication). However, a follow up by GROND \citep{gre08} about two days after the burst detected a faint object at the position of the afterglow at R=24.2 mag \citep{kru10}. 
%Because of this faintness and its position in the sky, 
{\em Swift} terminated the follow-up 2.14 days after the trigger.%, when the count rate in the 0.3-10 keV reached a level of 10$^{-3}$ counts s$^{-1}$.

\section{Data reduction and analysis}
\label{sec_analyze}

%\subsection{Prompt data}

\subsection{X-ray data}

All the XRT observations performed for this burst are quoted in Table \ref{tab:tab0}.  The X-ray light curve, taken from the XRT repository \citep[see][]{eva07}, shows the typical steep-flat-steep decay, as observed in about 70$\%$ of the Swift GRBs \citep[e.g.][]{lia09}.  A double broken power law model provides an excellent fit to the X-ray light curve ($\chi^2_\nu=0.89$ with 53 degrees of freedom), with best fit decay indexes $\alpha_{X,1}=2.3\pm0.9$, $\alpha_{X,2}=-0.04\pm0.02$ and $\alpha_{X,3}=1.34\pm0.07$ and break times at $t_{break,1}=97^{+9}_{-23}$ s and $t_{break,2}=925^{+150}_{-115}$ s after the trigger.

The hardness ratio between the 1.5$-$10 keV band and the 0.3$-$1.5 keV band shows some variability with time, in particular during the plateau, where no spectral variation is typically observed. However, a time resolved spectral analysis in the temporal ranges 100-300 s, 300-700 s, 4.0-10.0 ks and 10.0-98.9 ks after the trigger, shows that this behavior is not statistically significant. The average spectrum is well fitted by an absorbed power law. Assuming two absorption components, one fixed at the Galactic value of $N_H=6.5\times 10^{20}$ cm$^{-2}$ \citep{kal05} and the second free to vary, the best fit parameters are $\beta=1.0\pm0.1$ and an intrinsic absorption of $N_{H,host}=(4.0^{+2.8}_{-3.1})\times 10^{20}$ cm$^{-2}$.

\begin{table}
\begin{center}
\caption{Swift/XRT observations of GRB 101024A: from left to right are the Swift/XRT sequence (observation) identifying number, the start and end date of each sequence in seconds from the burst trigger time.\label{tab:tab0}}
\begin{tabular}{lcc}
\hline
Swift/XRT   & Start time &   End time \\
sequence ID & (s)        &  (s)       \\
\hline
00437016000 &  95.1   &   22 326 \\
00437016001 &  34 715 &   69 984 \\
00437016002 &  97 113 &  115 689\\
00437016003 & 143 337 &  185 155\\
\hline
\end{tabular}
\end{center}
\end{table}

\subsection{Zadko optical data}
\label{obs_opt}

\begin{table*}
\caption{Observations of \object{GRB 101024A} in optical.}             % title of Table
\label{log_obs}      % is used to refer this table in the text
\centering                          % used for centering table
\begin{tabular}{ccccc}        % centered columns (4 columns)
\hline\hline                 % inserts double horizontal lines
Date since burst & Exposure time & R magnitude    & Telescope & reference \\    % table heading 
(s)              &   (s)         &                &           &           \\          
\hline                        % inserts single horizontal line
218.2            & 6.4           & 16.6 $\pm$ 0.3 & Zadko     & This work \\
224.6            & 6.4           & 16.7 $\pm$ 0.3 & Zadko     & This work \\
232.1            & 8.6           & 16.9 $\pm$ 0.3 & Zadko     & This work \\
240.7            & 8.6           & 17.3 $\pm$ 0.3 & Zadko     & This work \\
274              & 30            & 17.6 $\pm$ 0.3 & Zadko     & This work \\
320              & 30            & 17.8 $\pm$ 0.3 & Zadko     & This work \\
364              & 30            & 18.0 $\pm$ 0.3 & Zadko     & This work \\
409              & 30            & 18.0 $\pm$ 0.3 & Zadko     & This work \\
1416             & 600           & 18.7 $\pm$ 0.4 & GRAS06    & \citet{hen10} \& Hentunen (priv. com.) \\
2074             & 600           & 20.2 $\pm$ 0.8 & GRAS06    & \citet{hen10} \& Hentunen (priv. com.) \\
4145						 & 720           & 19.5	$\pm$ 0.3 & AAVSO node& This work \\
4942						 & 720           & 19.9	$\pm$ 0.4 & AAVSO node& This work \\
5740						 & 720           & 20.1	$\pm$ 0.4 & AAVSO node& This work \\
6467						 & 600           & 21.0	$\pm$ 0.8 & AAVSO node& This work \\
10456						 & 5400          & 21.4	$\pm$ 0.5 & AAVSO node& This work \\
160440           & 3000          & 24.2 $\pm$ 0.3$^{\mathrm{a}}$ & GROND     & \citet{kru10} \\
\hline                                   %inserts single line
\end{tabular}
\begin{list}{}{}
\item[$^{\mathrm{a}}$] We converted the GROND r' magnitude expressed in the AB system into the standard R magnitude expressed in the Vega system.
\end{list}
\end{table*}

GRB 101024A was observed near the south celestial pole, and occurred at a time when only Australia was able to observe it. As such, few optical data are available. We have undertook a complete re-analysis of the Zadko data announced in \citet{laa10b} and the AAVSO data reported by \citet{sto10}, in order to obtain the best confidence photometric values reported in this paper. We performed aperture photometry using the AudeLa software\footnote{http://www.audela.org} \citep{klo08}, and list in Table \ref{log_obs} the photometry results. The large uncertainties (0.3 mag) are due to the reference star uncertainties extracted from the NOMAD1 catalog. %For completeness, we also report the relative photometry (i.e. without systematics) in Table \ref{phot_relative}.

For completeness, we have also retrieved from GCN circulars all available optical observations of this burst, and list them in Table \ref{log_obs} together with the Zadko telescope observations. Values reported from GRAS06 telescope have been checked by the GRAS06 team (Hentunen 2010, private communication). Because of the afterglow faintness, the UVOT on board Swift reported only upper limits in the temporal range used in this work, and are not reported.

\begin{figure}
   \centering
   \includegraphics[width=9cm]{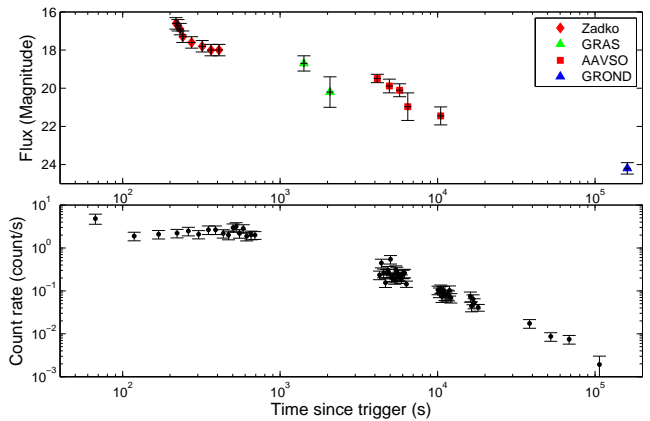}
      \caption{Light curve of \object{GRB 101024A} in optical (top panel) and X-ray (bottom panel). 
              }
         \label{lc_all}
   \end{figure}

We plot the optical light curve in Fig. \ref{lc_all}. As can be seen, the initial part of the optical light curve is quite similar to the canonical {\em Swift} X-ray light curve, i.e. a fast decay followed by a plateau, and a late steepening.
however, the optical late part is quite strange. Indeed, the GROND observation is not compatible with the decay rate measured from the AAVSO observations. These observations, in turn, are not compatible with the GRAS06 ones. This could be due to an optical flare between $\sim 2000$ and 4000 seconds after the burst. However, the GROND data are also not compatible with the decay rate measured from the GRAS06 data. This last deviation is not very large, and could be due either to a fluctuation of the optical afterglow or to a contribution of the host galaxy. Because of the lack of follow-up, we cannot conclude if this late GROND observation is the afterglow or rather the host galaxy. 

We tried to fit the light curve using broken power laws. If the early ($<\sim 1500$s) optical light curve can be described as a broken power law with best fit parameters of $\alpha_{O,1}=5.3\pm0.1$, $t_{b,1}=263\pm9$ s, $\alpha_{O,2}=0.6\pm0.1$, the late decay cannot be fitted with these models (the best $\chi^2_\nu$ we obtained was 6.8, allowing one more break). This large value of $\chi^2_\nu$ together with the residual analysis make us conclude that the usual power-law decay is not applicable to this afterglow, even when removing the last GROND observation (assumed to be affected by the host galaxy luminosity), and that another component is at play within the optical data.

%Excluding the GROND data, we fit the light curve using a double broken power law model. The very early part of the light curve has  $t_{b,2}=1500\pm100$s and $\alpha_{O,3}=4.0\pm1.0$. In such a case, the decay rate after the plateau phase is extreme and at least a factor of 2 higher than the most steep post-plateau decay rate typically observed. On the other hand, considering the late time observations of GROND as the afterglow flux level at that time, the best fit decay rate is more consistent with typical values observed in the majority of the afterglows: we obtain in this case that a broken power law model best fit the data, with a final decay rate after the first break ($t_{b,1}$) of $\alpha_{O,2}=1.1\pm0.1$. However is such a case the first GRAS06 observation lies about 3 sigma above the model prediction.

\begin{figure}
   \centering
   \includegraphics[width=7.5cm]{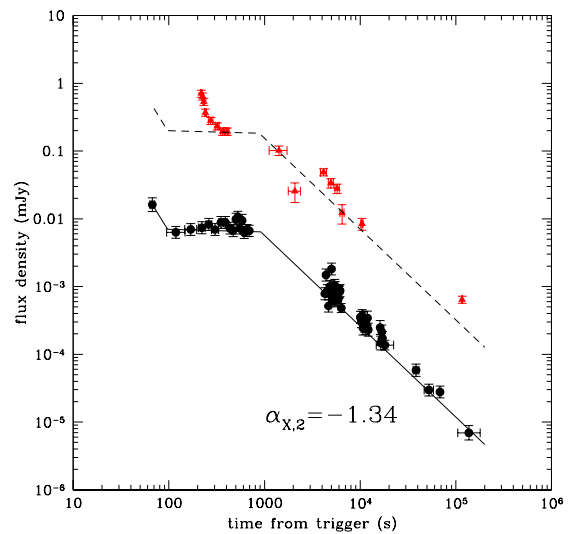}
      \caption{The X-ray best fit light curve (in black) compared to the optical data points (in red). See the electronic edition for colors.
              }
         \label{fig2}
   \end{figure}

Finally, we note that the last Zadko data features a steady flux, compatible with what is observed in X-rays. Applying the X-ray best fit light curve model to the optical data we find that both the plateau decay index and the epoch at the end of the observations ($t_{break,2}$) are consistent with the optical data (see Fig. \ref{fig2}). However, the post-plateau decay would over predict the second GRAS06 point of about 4 sigma while the late time GROND data would be $\sim3$ sigma above the predicted flux (consistent with the presence of a host galaxy), indicating again the need of another optical component. In addition, in order to reproduce the early optical light curve, a Gaussian (optical peak: 157 s after the trigger; peak width: 50 s) has to be superimposed on the X-ray best fit broken power law model.

%\begin{table}
%\caption{Zadko relative photometry of \object{GRB 101024A}.}             % title of Table
%\label{phot_relative}      % is used to refer this table in the text
%\centering                          % used for centering table
%\begin{tabular}{ccccc}        % centered columns (4 columns)
%\hline\hline                 % inserts double horizontal lines
%Date since  & Exposure  & Relative        & Relative  \\    % table heading 
%burst       & time      & magnitude       & magnitude \\
%(s)              &   (s)         & (reference star)& (GRB)      \\          
%\hline                        % inserts single horizontal line
%274              & 30            & -10.830 $\pm$ 0.047 & -9.414 $\pm$  0.047      \\
%320              & 30            & -10.782 $\pm$ 0.025 & -9.138 $\pm$  0.003      \\
%364              & 30            & -10.806 $\pm$ 0.006 & -8.965 $\pm$  0.147      \\
%409              & 30            & -10.812 $\pm$ 0.006 & -8.993 $\pm$  0.060      \\
%\hline                                   %inserts single line
%\end{tabular}
%\end{table}

%\subsection{Estimation of the redshift}

%Bruce for estimation of the redshift using the Amati relation, the B\&G relation, and if possible the pseudoZ of JLA. IN CASE, DISCUSS ALSO THE UVOT CONSTRAINTS...

\section{The early optical afterglow}
\label{sec_sample}

Zadko observations show an early optical steep decay, starting 218 seconds after the burst up to about 230 seconds. The steepness smooths to a shallower decay. While the plateau decay rate and the following steepening epoch are consistent with that measured from the X-ray light curve, the early optical  behavior is not tracked in the X-ray, suggesting a separate component.

The early optical light curve of 101024A may be interpreted as reverse shock afterglow emission. However, the initial decay index $\alpha_{O,1}$ is much higher than the expected value \citep[$\sim2$][see also Corsi et al. in preparation]{kob00}. We note however that a shift in the $T_0$ value may reduce this large decay index \citep[see e.g.][]{laz06}.

In some GRBs, the high energy prompt emission revealed an optical counterpart. We investigate whether this may be the case for GRB 101024A. The high energy (15-350 keV) prompt emission ended about 20 seconds after the trigger, and about 200 s before the optical observation. Such a large temporal lag between high energy and optical emission do not support this interpretation.

The steep decay during the early part of the optical light curve, together with the global behavior of the late optical data may also suggest the presence of a flaring activity. Other GRBs showed such features (e.g. GRB 030329, GRB 021004). GROND observations (e.g. GRB 081029, Nardini et al. in preparation) have recently shown optical strong flaring activity (or re-brightening), superimposed on the afterglow decay \citep[see also][]{gre11}. It has been proposed that ISM density enhancement of a factor of about 10 may produce early optical flares or re-brightening; a different origin than the one typically invoked for X-ray afterglows \citep[i.e. internal shocks][]{laz02}. In such a case, the absence of simultaneous flares observed in X-ray is due to the fact that above the cooling frequency the radiation mechanism is not sensitive to the surrounding medium density.

We used the closure relations between the decay and spectral indices applying to the late afterglow \citep{sar98, sar99} to constrain the fireball regime. The equation $\alpha - 1.5 \beta =  0.0$ is in agreement within errors with the values of $\alpha_{X,3}$ and $\beta$. This indicates that for GRB 101024A the X-ray band is below the cooling frequency if the surrounding medium is an InterStellar Medium (ISM) and that $p=2.8\pm0.1$. No other cases of the standard model apply for this burst \citep[see][for a list of closure relations tested on GRB 101024A]{gen06}. In such a case, the X-ray band should be sensitive to the clumpy medium and reproduce the optical flares. Moreover, because the cooling frequency decreases with time in case of an ISM, this consideration should be true also during the early afterglow. Despite the fact that the late X-ray light curve presents small fluctuations that may trace the small flares seen at the same time in optical, this is not the case for the initial steep decay. Thus, the clumpy medium interpretation does not apply to the initial data of GRB 101024A.

\section{The plateau}
\label{sec_discu}

\subsection{Standard model}

In the standard model, it is possible to observe a rising part in the afterglow if the injection frequency has not crossed the observation band. In such a case, the decay index should be 1/6 \citep{sar98}. The value we obtained in the X-ray light curve during the plateau for $\alpha_{X,2}$ is compatible with this value within 3 sigma. Once the injection frequency has crossed the observation band, we should observe a decay index between 0.7 and 1.3 (depending on the cooling regime and the fireball parameters), again compatible with the value of $\alpha_{X,3}$. However, in such a case, we should also observe some spectral variation, which is not observed in the case of this burst. Therefore, we can reject the hypothesis that the plateau phase is due to the standard model normal evolution.

\subsection{Energy injection}

It has been proposed by several authors \citep[e.g.][]{pan06} to account for the plateau phase by injecting energy into the fireball. We recall here that the afterglow is emitted by a population of accelerated electrons that radiate by synchrotron effect into a magnetic field produced by the fireball itself. Accordingly, there are 3 methods to inject energy: by inserting some energy into the fireball (by shocks), by transferring some energy not used of the fireball into the electron (modifying the $\epsilon_e$ parameter), and by increasing the magnetic field of the fireball (modifying the $\epsilon_B$ parameter).

The observed flux can be expressed as:

\begin{equation}
F_\nu \propto E^{-\delta} \epsilon_e^{-\omega} \epsilon_B^{-\Lambda} t^{-\alpha}
\end{equation}

If one (or a combination) of $E^\delta$, $\epsilon_e^\omega$, or $\epsilon_B^\Lambda$ is proportional to $t^{+\alpha}$, then the light curve will feature a plateau. The difficulty of this model is that the modification of the temporal decay does not depend on the wavelength, i.e. assuming that the optical and X-ray bands are in the same cooling regime the plateau is observed at both wavelengths. In addition, if a specific frequency (e.g. the cooling frequency) lies between the optical and X-ray bands, then the optical band decay will not be identical to the X-ray one. However, at the end of the energy injection (i.e. at the end of the X-ray plateau) the optical light curve will feature a break. In several cases where energy injection is claimed, the presence of this break is not checked in available data, or even ruled out by the observations, see e.g. \citet{xin10}, \citet{cus05}, or \citet{nou05}. The simultaneity of the break can be relaxed if one removes one of the main hypotheses of the fireball model, that the electron population is distributed according to a single power law. 

In the case of GRB 101024A however, as we noted in Sec. \ref{obs_opt}, if we assume that the initial steep decay is not linked to the afterglow, the data cannot rule out a simultaneity of the breaks in optical and in X-ray, as it is expected in case of energy injection. However, the optical data are too scarce to confirm the energy injection as the only explanation for the plateau phase, and we conclude that an energy injection during the plateau phase is not excluded by the data and may explain the plateau phase.

\subsection{Gravitational waves}

The shallow decay phase in the early X-ray afterglows of gamma-ray bursts may be the signature of the formation of a highly magnetized millisecond pulsar, pumping energy into the fireball on timescales longer than the prompt emission. This scenario has also led to the more speculative idea of the nascent neutron star undergoing rotational instabilities, so called `bar modes'. A neutron star undergoing a bar mode instability would be a strong gravitational wave source. \citet{cor10} argued that the GRB early afterglow plateau could be a `smoking gun' for gravitational wave emission and a target for Advanced LIGO and Virgo. 

Assuming that GRB 101024A follows this evolutionary scenario, could it have been detected by Advanced LIGO? Given the expected gravitational wave luminosity from a bar node instability, and the sensitivity of Advanced LIGO, it would probably not be detectable unless GRB 101024A occurred in a host galaxy at a distance less than about 150 Mpc. Nonetheless, it is an exciting prospect that the potential exists for probing the strong gravity regime of compact objects in both the electromagnetic and gravitational wave spectra.

\section{Conclusions}
\label{sec_conclu}

We present observations of GRB 101024A obtained with the Zadko telescope in optical and with {\em Swift} at high energy. In the optical, the afterglow data reveal a complicated behavior. The light curve first decay with a large and unusual decay index, that cannot be understood in the standard framework of the fireball model. After this strong decay, the light curve features a flattening. The simultaneously observed X-ray data feature a plateau that could indicate 
late energy injection. The optical data do not rule out this hypothesis. The optical light curve features also an erratic behavior with late re-brightening, with no clear explanation. Finally, we note that the last observation of this burst made by GROND has not a clear explanation. It could be either another very late flare or the consequence of the host galaxy luminosity. A further deep photometric observation on a 8m class telescope could resolve this mystery.

The GRB 101024A optical data are too sparse to provide a complete analysis of the burst properties. We note this highlights the problem of a lack of ground based follow-up of GRB afterglows. If a burst commenced during the European night, it can be continuously followed by small robotic telescopes in Europe, medium size instrument in the Canary Islands, large facilities in the USA, and Chile, over 12 hours. Conversely, a burst occurring when the night is almost finished in South America will be poorly sampled by the Australia and Japan facilities. This geographic bias could be helped by the robotisation of several 2-4 meter class telescopes at Australian and African longitudes.

\begin{acknowledgements}
Thanks to V.P. Hentunen who kindly provided the updated and verified data of the GRAS06 telescope. This work made use of data supplied by the UK Swift Science Data Centre at the University of Leicester. This work has been financially supported by the GdR PCHE and the GDRE "Exploring the Dawn of the Universe with Gamma-Ray Bursts" in France, and from the Italian Space Agency (ASI). We finally thank an anonymous referee for his/her valuable comments.
\end{acknowledgements}

\end{document}